\documentclass[aip,cha,amsmath,amssymb,reprint]{revtex4-1}

\usepackage{graphicx}% Include figure files
\usepackage{dcolumn}% Align table columns on decimal point
\usepackage{bm}% bold math
\usepackage{color}
\definecolor{red}{rgb}{1,0,0}

\begin{document}
    
    % old short title: Heterogeneities in power grids strongly enhance non-Gaussian features of frequency fluctuations
%    \newcommand{\tit}{Heterogeneities in electricity grids strongly enhance non-Gaussian features of frequency fluctuations under stochastic power input}

\title[Heterogeneities in power grids strongly enhance non-Gaussian features of frequency fluctuations]{Heterogeneities in electricity grids strongly enhance non-Gaussian features of frequency fluctuations under stochastic power input}

\author{Matthias F. Wolff}
\email{mawolff@uos.de}
\affiliation{Universit\"at Osnabr\"uck, Fachbereich Physik,
    Barbarastra{\ss}e 7, 49076 Osnabr\"uck, Germany}

\author{Katrin Schmietendorf}
\affiliation{Center for Nonlinear Science, 
    Westf\"alische Wilhelms-Universit\"at M\"unster, 
    Correnstra{\ss}e 2, 48149 M\"unster, Germany}

\author{Pedro G. Lind}
\affiliation{Department of Computer Science,
OsloMet - Oslo Metropolitan University,
Pilestredet 35, PS428,
0166 Oslo, Norway}

\author{Oliver Kamps}
\affiliation{Center for Nonlinear Science,
    Westf\"alische Wilhelms-Universit\"at M\"unster, 
    Correnstra{\ss}e 2, 48149 M\"unster, Germany}
    
\author{Joachim Peinke}
\affiliation{Universit\"at Oldenburg, {Institut f\"ur Physik \& ForWind},
    K\"upkersweg 70, 26129 Oldenburg, Germany}

\author{Philipp Maass}
\email{maass@uos.de}
\affiliation{Universit\"at Osnabr\"uck, Fachbereich Physik,
   Barbarastra{\ss}e 7, 49076 Osnabr\"uck, Germany}

\date{\today}

\begin{abstract}
Stochastic feed-in of fluctuating renewable energies is steadily increasing in modern electricity grids
and this becomes an important risk factor for maintaining power grid stability. Here we study the impact of
wind power feed-in on the short-term frequency fluctuations in power grids based on
an IEEE test grid structure, the swing equation for the dynamics of voltage phase angles, and a
series of measured wind speed data. External control measures are accounted for
by adjusting the grid state to the average power feed-in on a time scale of one minute.
The wind power is injected at a single node by replacing one of the conventional generator nodes
in the test grid by a wind farm. We determine histograms of local frequencies
for a large number of one-minute wind speed sequences taken from the measured data and for
different injection nodes. These histograms exhibit a common type of shape, which can be described
by a Gaussian distribution for small frequencies and a nearly exponentially decaying tail part.
Non-Gaussian features become particularly pronounced for wind power injection
at locations, which are weakly connected to the main grid structure. This effect is only present
when taking into account the heterogeneities in transmission line and node properties of the grid,
while it disappears upon homogenizing of these features.
The standard deviation of the frequency fluctuations increases linearly with the average injected 
wind power.
\end{abstract}

\maketitle

\begin{quotation}
Electric energy supply is of utmost importance for
industrial companies and private households, and it will
become even more relevant in connection with
actions taken for mitigating climate change effects.
The increasing feed-in of wind and solar power into electricity 
grids poses new challenges for maintaining their stability.
% External control measures currently
%act on scales of about one minute. Below this time scale,
%power grids are essentially autonomous, with only
%simple automatic regulations. 
The stochastic nature of wind speeds
and solar irradiation yield short-term fluctuations of the local
frequencies with possible large deviations from the nominal frequency of
the desired synchronous operating state. These large deviations can
form nuclei for grid instabilities, which can range from single line overloads
to malfunction of larger grid parts. Short-term frequency stability
therefore must be assessed under erratic power feed-in. 
Here we study
this problem for wind power injection. 
We find that wind power feed-in leads to
exponentially decaying tails of local frequency
distributions. This implies rare large frequency
fluctuations to occur much more frequent than expected from assuming simple
Gaussian statistics. A decisive factor for the appearance of the
non-Gaussian large frequency fluctuations are the
heterogeneities in transmission line and node properties.
The non-Gaussian features are much less significant and almost negligible
in a homogenized grid.

\end{quotation}

\section{INTRODUCTION}
\label{sec:introduction}
The steadily increasing share of fluctuating wind and solar power in electricity grids
raises new questions on the assessment and control of grid stability. To tackle this problem,
different aspects and challenges need to be considered and mastered. One aspect 
is the grid topology, which can be generated
artificially by some reasonably developed algorithm, which then allows one to perform an ensemble averaging.
\cite{Menck/etal:2014, Schultz/etal:2014, Jung/Kettemann:2016, Auer/etal:2016, Auer/etal:2017} Another option is to
use test grids like those provided by the Institute of Electrical and Electronics Engineers (IEEE).\cite{Schiel/etal:2017, Wolff/etal:2018} 
It is also possible to analyze specific motifs in a grid.\cite{Schultz/etal:2014a, Kim/etal:2018}
Or one can try to use real grid
structures, which unfortunately are in general not provided by the
network operating companies. However, there are a number of initiatives, such as open\_eGo\cite{Mueller/etal:2018}, SciGrid\cite{SciGRIDv0.2} and others, which try to obtain real
grid structures, based mainly on information taken from Open Street Map. Characteristic features of the grid
structure are different for
different voltage levels and a further issue is the modeling between these voltage levels.

Another aspect is the modeling depth, that means the question whether one can rely on a simple quasi-
stationary approach
based on power flow equations,\cite{Anghel/etal:2007, Chertkov/etal:2011, Schiel/etal:2017} or whether one needs
to couple these flow equations to the voltage angle dynamics described by the
swing equation with possible further extensions for including dynamics of voltage amplitudes.\cite{Schmietendorf/etal:2014, Auer/etal:2016} In addition there exist different
models to describe generator and load nodes,\cite{Nishikawa/Motter:2015} and it seems to be relevant to take into
account the impact of reactances in the
coupling of loads and generators to the grid.\cite{Wolff/etal:2018}

The necessary modeling depth for obtaining reliable results will depend also on the
relation of several time scales, such as scales for primary and secondary control, intrinsic dynamical scales and the scales associated
with the fluctuations of renewable energy sources.\cite{Schmietendorf/etal:2017, Auer/etal:2017, Haehne/etal:2018, Haehne/etal:2019, Schaefer/etal:2018} To account for these fluctuations, we need good descriptions of the stochastic
dynamics of wind and solar power, which involves features coming from atmospheric turbulence, cloud effects and questions related to
how specific engineering setups affect the transfer of a physical source, e.g.\ wind speed, to the injected power.

In quasi-stationary approaches, the focus is generally on how power flows along transmission lines are modified due to a change in
renewable power generation and whether weak points can be identified, where lines become overloaded with high probability.\cite{Albert/etal:2004, Anghel/etal:2007, Wang/Rong:2009, Hines/etal:2010,  Chertkov/etal:2011, Andrychowicz:2016, Schiel/etal:2017}
Fewer studies are concerned yet with the modified power flow dynamics caused by fluctuations of wind and solar power on short time
scales.\cite{Schmietendorf/etal:2017, Auer/etal:2017, Schmietendorf/etal:2018}
These time scales have to be put in relation to those of external control measures.
%Here in particular time scales are of interest that are below those of the external measures of secondary control, i.e.\
%times below one minute and less.
Sudden large deviations of local frequencies from the nominal value can form nuclei for grid
instabilities, which can range from single line overloads to malfunction of larger grid parts, up to cascading failures spanning
large fractions of the whole grid. A better understanding of the statistics of local frequency fluctuations under the stochastic
input of wind and solar power is needed to develop reliable risk estimates of grid failures and strategies to balance risk factors
with investments in higher grid stability. Another aspect of the fluctuating power input is a possible reduction of frequency quality, i.e.\ the percentage of time where the grid
operates in a given frequency range. 
This can be estimated from distributions of local frequencies.

A challenge in treating short-time dynamics in power grids is how to take into account
the effect of the external control measures on longer times scales. For example, in a time period of high average wind speed, the
conventional generators in the grid will generate less power than in a period of low average wind speed. This means that the state
of conventional generator nodes depends on the average wind speed or wind power level. We introduce a
concept in this study, where the grid state is adapted to the average wind power level.

\begin{figure*}[t!]
\includegraphics[width=\textwidth]{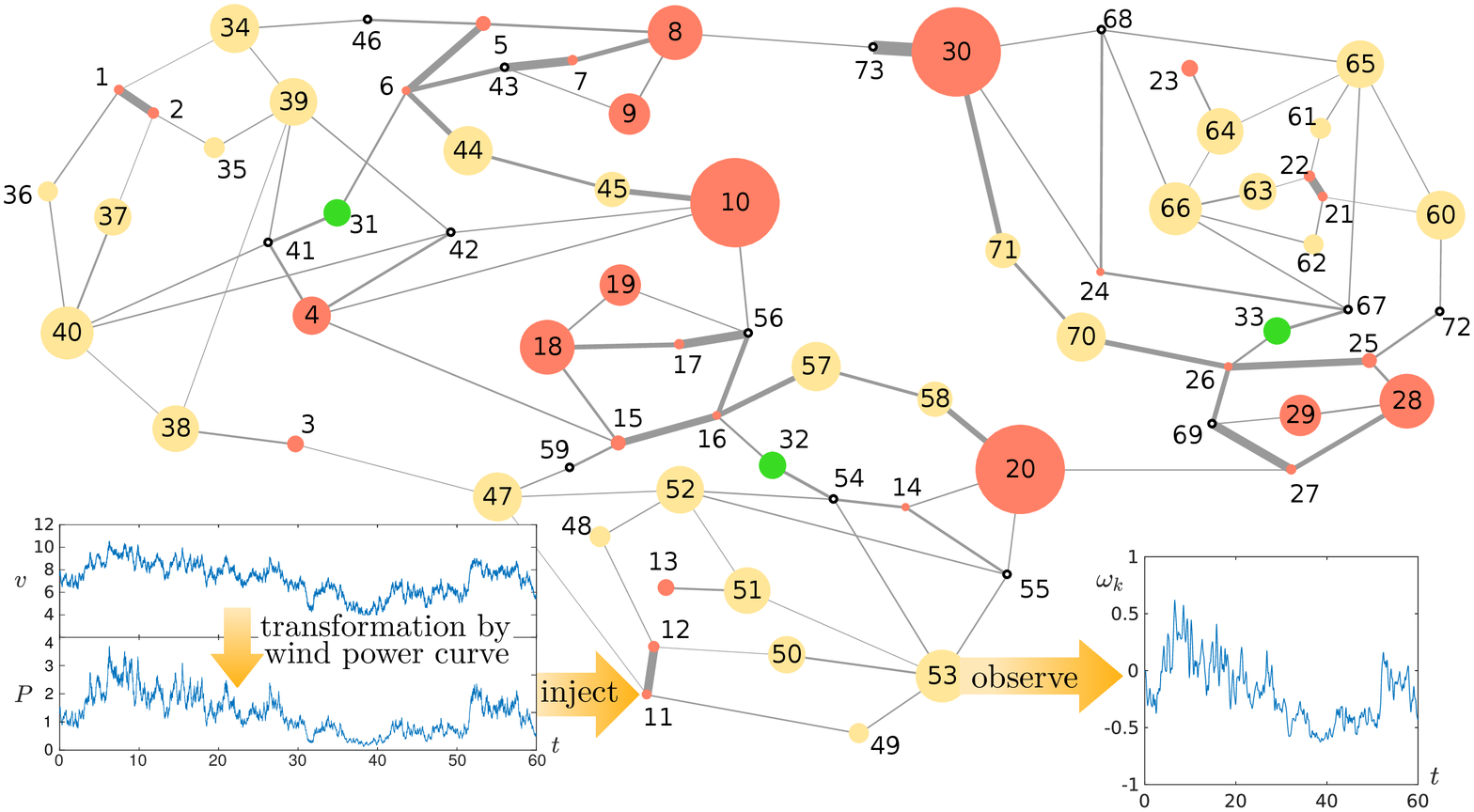}
\caption{Sketch of the IEEE RTS-96, consisting of 30 generators labeled from 1 to 30 (red circles), 3 synchronous condensers labeled 31, 32, 33 (green circles), and 40 load nodes labeled from 34 to 73 (yellow circles/open white circles). The small white circles refer to load nodes with vanishing mechanical power. The synchronous condensers have a fixed size, and the size of other symbols has been scaled proportional to their mechanical power. The nodes are connected by 108 transmission lines, where the thickness of the lines marks the strength (modulus) of the respective complex admittances.
    %, which is defined by $ [|(Y_{0})_{jk} | + |(Y_{0})_{kj} |]/2 $, where $ (Y_{0})_{jk} $ are the elements of the admittance matrix.
The insets illustrate the injection of wind power
at node 11 due to fluctuating wind speeds $v(t)$ and resulting fluctuations of the frequency $\omega_k$ at node $k=53$.
\label{fig:model-illustration}}
\end{figure*}

Specifically, we focus on the distribution of short-time local frequency fluctuations in the
IEEE Reliability Test System 1996 (IEEE RTS-96)\cite{Grigg/etal:1999}
if one of its
conventional generator nodes is replaced by a node with power feed-in from a wind farm, see Fig.~\ref{fig:model-illustration}.
For this feed-in,
we take data of wind velocities measured at a tower in the North Sea with a sampling rate of 1~Hz.
%\cite{FINOdata}
\footnote{FINO I project and database (2016). The FINO project is supported by the German Government through BMWi and PTJ. Available at \texttt{http://www.bsh.de}}
Both the generator and load nodes of the IEEE RTS-96
are described by the synchronous machine model. The
voltage angle dynamics is determined by the swing equation with forcing by the imbalance of
mechanical and electric power. The latter is given by the nonlinear power flow equations, where voltage magnitudes
are considered to be fixed.

We investigate also how the distributions of local frequency fluctuations change if
transmission line and node heterogeneities in the IEEE RTS-96 are homogenized by an averaging procedure.
Homogenized grid properties are often used in
simplified modelings, because they reduce the computational effort for solving the nonlinear dynamical equations,
and make it easier to find fixed point solutions and to maintain numerical stability.  However, the use of homogenized
grid properties may lead to an underestimation of failure probabilities and a wrong identification of weak parts in the grid.
\cite{Wolff/etal:2018}

As a key result of our study we find that the distributions of local frequency exhibit non-Gaussian
features due to tails with approximately exponential decay. These non-Gaussian
features can be very pronounced at certain grid nodes.
For small frequency fluctuations, the distributions have a nearly Gaussian shape. This Gaussian
core part of the distributions gives essentially the fluctuation width, i.e.\ the standard deviation of frequency fluctuations,
while the nearly exponential tails are important for estimating probabilities of rare large fluctuations.
In the homogenized grids, the non-Gaussian features are less pronounced and the fluctuation widths are much smaller.

The paper is organized as follows. In Sec.~\ref{subsec:modeling-dynamics} we introduce the basic dynamical equations and the essential features of the IEEE RTS-96 structure
relevant for this study. In Sec.~\ref{subsec:modeling-stochastic-generation} we describe the stochastic wind power generation based on the measured wind speed data,
and in Sec.~\ref{subsec:modeling-feed-in} we discuss the wind power feed-in.
Our results for the local frequency distributions are detailed in Sec.~\ref{sec:local-frequency-distributions}.
Section~\ref{sec:conclusions} closes our work with a discussion of the implications of our findings and an outlook for further research.

\section{POWER FLOW, STOCHASTIC INPUT AND GRID STRUCTURE}
\label{sec:modeling}

\subsection{Power flow dynamics}
\label{subsec:modeling-dynamics}

We model the power flow dynamics based on the widely\cite{Rohden/etal:2012, Menck/etal:2014, Jung/Kettemann:2016, Schaefer/etal:2018:2, Auer/etal:2016, Auer/etal:2017, Tamrakar/etal:2018} used synchronous machine model for
loads and generators.\cite{Machowski/etal:2008, Filatrella/etal:2008, Nishikawa/Motter:2015}
 With the nominal frequency $f_{\rm r}$ (50~Hz in Europe) and corresponding
angular frequency $\omega_{\rm r}=2\pi f_{\rm r}$,
the voltage at node $j$ is $V_j(t)=|V_j|\,{\rm Re}\,[e^{-i\omega_{\rm r} t+i\theta_j(t)}]$, where
$\theta_j(t)$ is
the phase angle describing the deviation from the synchronous state of operation;
the moduli $|V_j|$ of the voltages are considered to be time-independent.
The imbalance $(P_j^{\rm (m)}-P_j )$
between the ``mechanical'' powers $P_j^{\rm (m)}$
and electrical powers $P_j$ drives the phase angle $ \theta_{j} $ according to the swing equation
\begin{eqnarray}
H_j \ddot{\theta}_j + D_j \dot{\theta}_j &=& P_j^{\rm (m)}-P_j \cr
& & \cr
&=&
P_j^{\rm (m)}- \sum_k K_{jk}\sin\left(\theta_j-\theta_k-\gamma_{jk}\right).
\label{eq:swing}
\end{eqnarray}
Here the 
%electric power is expressed by the voltages
%by eliminating the currents with the aid of Kirchhoff's equations, resulting in
coupling constants are $K_{jk}=|V_j||V_k||Y_{jk}|$, where $Y_{jk}=|Y_{jk}|\exp[i(\gamma_{jk}+\pi/2)]$
are the elements of the grid admittance matrix $Y$. The $H_j$ are inertia constants
of the synchronous machines, i.e.\
connected with the rotating mass of a conventional generator or motor. The damping
constants $D_j$ effectively 
account for 
%both electrodynamic and mechanical damping effects, as well as 
primary control measures,
%\cite{ucte}
\footnote{See \texttt{https://www.entsoe.eu} for ``UCTE Operation Handbook'' (2004).}

which drives the grid into a synchronous state of operation.
From a mechanical perspective, Eqs.~\eqref{eq:swing}
correspond to a Newtonian dynamics of nonlinearly coupled oscillators with
damping. These are often referred to as the ``second-order Kuramoto equations''.
\cite{Kuramoto1975, Acebron/etal:2005, Rodrigues/etal:2016}

We have solved the swing equations \eqref{eq:swing} numerically by applying a Runge-Kutta
solver of fourth order with a time step of $5\times 10^{-4}$~s. As for the parameters,
listed values for the IEEE RTS-96 were used, and estimates for the $H_j$, $D_j$ based on the findings reported in Ref.~\onlinecite{Motter/etal:2013}. For comparison with a simplified homogenized grid
variant, arithmetic means of line admittances as well as of consumed and generated powers
are taken. A detailed description of all parameters
is given in Ref.~\onlinecite{Wolff/etal:2018}.

If the mechanical powers $P_j^{\rm (m)}$ do not fluctuate, a
stationary synchronous state of the grid develops after some transient time,
where all frequency deviations $\omega_j=\dot\theta_j$ from the nominal
frequency $\omega_{\rm r}$ are zero and all $\theta_j$ are constant.
In principle, knowing the $P_j^{\rm (m)}$ and $K_{jk}=|V_j||V_k||Y_{jk}|$, the phase angles in this state
can be calculated by setting the left-hand side of Eq.~\eqref{eq:swing} to zero. However, for load nodes
the voltages $|V_j|$ are generally not known but the reactive powers $Q_j^{\rm (m)}$.
For determining the synchronous state, we thus have to solve the full power-flow equations
\begin{subequations}
\label{eq:pf}
\begin{eqnarray}
P_j^{\rm (m)} &=& P_j=\sum_{k} |V_{j}||V_{k}||Y_{jk}|
\sin\left(\theta_{j}-\theta_{k}-\gamma_{jk}\right),
\label{eq:pf-p}\\
Q_j^{\rm (m)} &=& Q_j=\sum_{k} |V_{j}||V_{k}||Y_{jk}|
\cos\left(\theta_{j}-\theta_{k}-\gamma_{jk}\right)\,,
\label{eq:pf-q}
\end{eqnarray}
\end{subequations}
which express the balance between mechanical and electrical powers, including Ohmic losses.
We solve these equations
by a Newton-Raphson method with starting angles
$(\theta_1,\dots,\theta_N)=(0,\dots,0)$, yielding a unique fixed point
vector $(0,\theta_2^\ast,\dots,\theta^{\ast}_N)$. Alternatively, one could
use the holomorphic embedding load flow method
for determining the fixed point of synchronous operation.\cite{Trias:2012}
If the mechanical powers are chosen to have values different from the ones listed for
the IEEE RTS-96, we have taken node 4 as the reference bus for determining the
fixed point state.

\subsection{Stochastic power generation}
\label{subsec:modeling-stochastic-generation}
In earlier times, where power was produced solely by conventional generators,
fluctuations of the $P_j^{\rm (m)}$ had to be considered for the load nodes.
Typically, the impact of corresponding load fluctuations can be
treated in a quasi-stationary approach based on the power-flow equations \eqref{eq:pf}. 
This is 
because significant changes of consumed power occur on time scales large compared to
relaxation times to the fixed point state, which lie in the range %of Eqs.~\eqref{eq:swing}
3-20 seconds.\cite{Wolff/etal:2018}

In the presence of stochastic feed-in from renewable energy sources, the impact of fluctuating $P_j^{\rm (m)}$ must be considered also for generator nodes.
The dynamics described by Eqs.~\eqref{eq:swing} can then no longer be ignored, because
power feed-in from wind and solar irradiation shows significant changes on
short time scales. To account for these fluctuations, corresponding stochastic processes for
$P_j^{\rm (m)}(t)$ need to be specified. When inserting these into
Eqs.~\eqref{eq:swing}, the phase angles $\theta_j(t)$ and
frequencies $\omega_j(t)=\dot\theta_j(t)$ become stochastic processes as well.

\begin{figure}[b]
\includegraphics[width=0.45\textwidth]{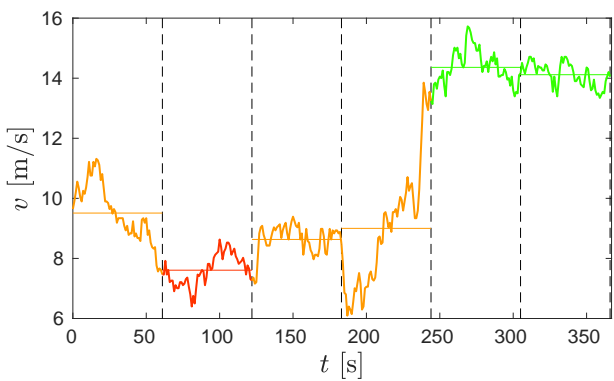}
\caption{Section of the wind speed series measured at a tower located in the North Sea
with a sampling rate of 1~Hz.$^{26}$ The vertical dashed lines indicate the division into
segments of one minute. Data are colored differently with respect to the mean wind speed in each segment
(horizontal lines), corresponding to a grouping into bins of widths 2~m/s, see the discussion in
Sec.~\ref{sec:local-frequency-distributions}.\label{fig:windparts}}
\end{figure}

In this paper we focus on the frequency fluctuations under stochastic feed-in of wind power.
An important statistical feature of wind speeds is that distributions of
velocity increments on even short times of order one second exhibit tails much heavier than
that of a Gaussian distribution due to the intermittent nature of turbulent flows.
Wind speed increments of, for example, 5~m/s are rare but
have probabilities that are by several orders higher than those expected from a Gaussian distribution
with mean and variance given by the measured data.
Recent results suggest that wind speed fluctuations are rather directly reflected in the generated power on short time scales of seconds and below.\cite{Haehne/etal:2018}
This would imply that rare
%, but still noticeably frequent
events of large sudden power changes could be an important
risk factor for maintaining grid stability.

To capture realistic features of the wind, we base our study
on a series of wind speeds $v$ measured at a tower located in the North Sea
with a sampling rate of 1~Hz.$^{26}$ From this series, we take
the data $\{v_n\}_{1\le n\le N}$ sampled in October 2016 ($N=2678400$) for our study. 
These wind speeds show strong fluctuations on all time scales.
The time window, where the grid state dynamics are
described by Eq.~\eqref{eq:swing} is, however, restricted by 
control measures. The synchronous state of operation
generated by primary control can have a frequency that deviates 
from the nominal value.
Secondary control measures tend to restore a synchronous state at the nominal
frequency. It involves time-delay feed-back and integration of power imbalances on time scales of about 30~seconds. We here
take this secondary control into account in an effective manner by
assuming that the ideal synchronous state at the nominal frequency is
restored after one minute.
Accordingly, we divide the series $\{v_n\}_{1\le n\le N}$ into
subsequent segments of one minute, yielding in total $N/60=44640$ sets, as illustrated
by the vertical dashed lines in Fig.~\ref{fig:windparts}.
These sets of one-minute data form the basis for the stochastic wind 
power input $P_j^{\rm (m)}(t)$ in Eq.~\eqref{eq:swing}.

\begin{figure}[b!]
\includegraphics[width=0.48\textwidth]{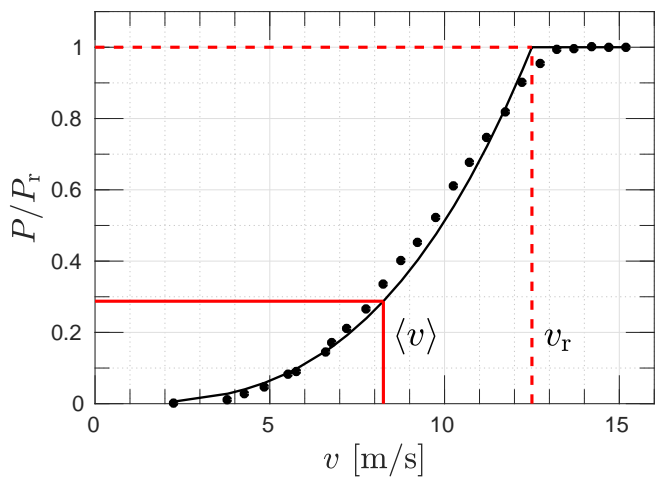}
\caption{Power curve giving the dependence of the wind power $P$ on the wind speed $v$. The power is normalized to the rated value $P_{\rm r}$, which is the saturation value for speeds larger than the rated
speed $v_{\rm r}$. The data points are extracted from Ref.~\onlinecite{Milan/etal:2014} and the solid black
line is a cubic fit to the data for $v<v_{\rm r}$. The solid red line indicates the mean value $\langle v\rangle$ of the
wind speeds sampled in October 2016 at the tower in the North Sea$^{26}$
and its corresponding power.
\label{fig:powercurve}}
\end{figure}

To convert the wind speeds into power, we make use of the so-called ``power curve'' $P(v)$,
which describes how wind speed translates to wind power on average. The power
curve $P(v)$ increases nearly as $\propto v^3$ for small $v$ up to a rated velocity $v_{\rm r}$, where
the power reaches its maximal allowed value $P_{\rm r}$ to prevent the wind turbines from getting
damaged (pitch control \cite{}), see Fig.~\ref{fig:powercurve}. The data points in this figure
are taken from Ref.~\onlinecite{Milan/etal:2014} and the solid line is a cubic fit to the data for small $v$ with a
cross-over to constant $P_{\rm r}$ for $v\ge v_{\rm r}$. The rated speed $v_{\rm r}=12.5$~m/s
was determined from the average $\langle v\rangle$ of the measured speeds $\{v_n\}_{1\le n\le N}$ and by
applying the rule of thumb $v_{\text{r}}=1.5\langle v \rangle$.\cite{Manwell:2009}
Let us note that the power curve has a ``cut-in speed'' at
about 4~m/s, and a ``cut-off speed'' at about 25~m/s, above which the blades of a wind turbine are turned away from the wind and
the power drops to zero. In our subsequent analysis we consider only sets of one-minute data with mean wind speeds
between 4~m/s and 18~m/s.

When substituting a conventional generator $j$ in the IEEE RTS-96 by a wind farm,
we also need to specify the size of the farm. This is done by demanding the power generated at the mean wind speed
$\langle v\rangle$ to be equal to the power $P^{\,\rm conv}_j$ of the substituted conventional generator. Accordingly, the
translation of wind speed $v$ into wind power $P^{\text{w}}_{j}(v)$ at node $j$ is
\begin{equation}
P^{\text{w}}_{j}(v)=\left\{\begin{array}{cl}
\displaystyle \frac{P^{\,\rm conv}_j}{\langle v \rangle^3}\, v^3\,, & v\le v_{\rm r}\,,\\[2.5ex]
(P_{\rm r})_j\,, & v\ge v_{\rm r}\,,
\end{array}\right.
\label{eq:powerfeedin}
 \end{equation}
with $(P_{\rm r})_j= P^{\,\rm conv}_j v_{\rm r}^3/\langle v \rangle^3$.

The measured wind speed data $\{v_n\}_{1\le n\le N}$ have a time resolution of one second. Much shorter time resolutions of
order one millisecond are necessary to integrate Eqs.~\eqref{eq:swing} with numerical accuracy. In order to
specify a time-continuous stochastic process for the feed-in of wind power in Eqs.~\eqref{eq:swing},
one could use a step-function approach, e.~g., by defining
$v(t)=v_1+\sum_{n=1}^{N-1} (v_{n+1}-v_n)\Theta(t-n)$
with $\Theta(.)$ the Heaviside jump function [$\Theta(x)=1$ for $x\ge0$ and zero otherwise; $t$ in units of seconds].
However, this approach would ignore fluctuations on shorter scales. We therefore prefer
to use a stochastic interpolation scheme between consecutive values $v_n$, $v_{n+1}$
that is explained in the Appendix.

\subsection{Wind power feed-in}
\label{subsec:modeling-feed-in}

As mentioned in the Introduction, it is important to take into account that the
external control measures are acting on longer time scales and lead to
a state of the controllable generators that is adapted to the average wind
power. Therefore, for a given set of one-minute data, we calculate the
mean wind power $\bar P^{\text{w}}_j$.
This average $\bar P^{\text{w}}_j$ is in general not equal to the
nominal power $P^{\,\rm conv}_j$ of the original conventional generators in the
IEEE RTS-96, implying a power imbalance between the total generated power
and the sum of the total consumed power $|P^{\,\rm load}_{\rm tot}|=-P^{\,\rm load}_{\rm tot}$ (assumed to be fixed)
and the Ohmic losses.
However,  due to the external control measures, we can view the conventional generators
to be uniformly scaled, $P^{\,\rm conv}_k\to\beta P^{\,\rm conv}_k$ so that the
total power generation $(\bar P^{\text{w}}_j+\beta\sum_{k\ne j}P^{\,\rm conv}_k)$
averaged over one minute remains the same as in the unmodified IEEE RTS-96.
The scale factor $\beta$ is
\begin{equation}
\beta=\frac{\sum_{l}P^{\,\rm conv}_l-\bar P^{\text{w}}_j}{\sum_{k\ne j}P^{\,\rm conv}_k}\,.
\end{equation}

Given $\bar P^{\text{w}}_j$ and the $\beta P^{\,\rm conv}_k$ for $k\ne j$, the correspondingly modified
IEEE RTS-96 assumes a new fixed point, which we determine as described in Sec.~\ref{subsec:modeling-dynamics}.
To solve Eqs.~\eqref{eq:swing} for a given one-minute realization of the feed-in process, we always start in this fixed point state,
thereby effectively taking into account the adaptation of the state due to external control measures.

\begin{figure}[t!]
    \includegraphics[width=0.45\textwidth]{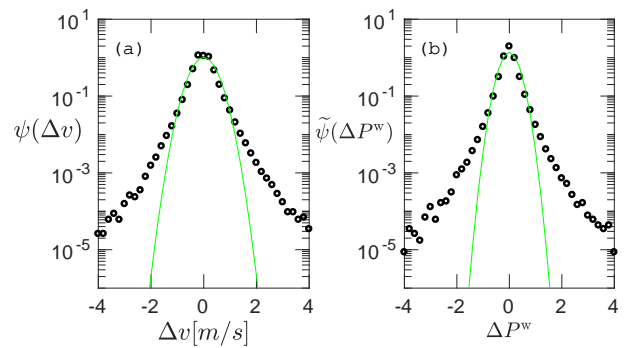}
    \caption{(a) Distribution $\psi(\Delta v)$ of one-second increments $\Delta v$ of 
    wind speeds obtained from the time series measured at a tower in the North Sea. In the sampling,
    only those data are included, where the average wind speed in a one-minute time interval
    lies in the range 8-10~m/s. (b) Corresponding distribution $\widetilde\psi(\Delta P^{\rm w})$ of increments
    of wind power feed-in
    at node $j=13$ that result from the transformation of wind speeds into powers described in Sec.~\ref{subsec:modeling-stochastic-generation}.
    The green lines in both graphs
    correspond to Gaussian distributions with zero mean and the same standard deviation a that of
     $\psi(\Delta v)$ and $\widetilde\psi(\Delta P^{\rm w})$. 
\label{fig:increments}}
\end{figure}

\begin{figure*}[t!]
    \includegraphics[width=0.9\textwidth]{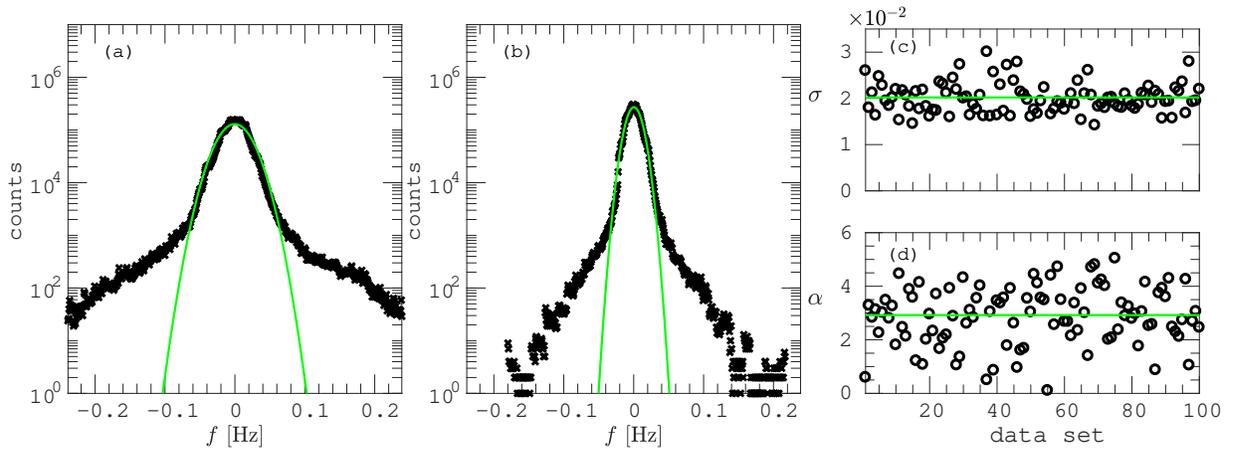}
    \caption{
            Example of local frequency histograms (black crosses)
            for a one-minute wind power feed-in $ P^{\rm w}(t) $ at node 13 whose mean wind speed  $ \bar{P}^{\rm w} $ belongs to the
            bin 8-10~m/s, (a) for the IEEE RTS-96 (after replacement of node 13 for wind-power feed-in),
            and (b) for the corresponding homogenized grid structure (as described in
            Sec.~\ref{subsec:modeling-dynamics}). Frequency data were sampled at all other nodes.
            The green lines correspond to Gaussian distributions with zero mean and standard deviation
            $\sigma$ calculated from the frequency fluctuations.
            The graphs (c) and (d) depict variations of $\sigma$ and the non-Gaussian parameter
            $\alpha$ [defined in Eq.~\eqref{eq:alpha}] for different one-minute power feed in at the same node $j=13$. The horizontal green lines show the average values. Non-Gaussian parameters for the homogeneous grid variant are much smaller (cf.\ Fig.~\ref{fig:alpha}).
        \label{fig:freqhist}}
\end{figure*}
%\section{IMPACT OF STOCHASTIC POWER INPUT}
\section{LOCAL FREQUENCY DISTRIBUTIONS}
\label{sec:local-frequency-distributions}
The numerical solutions of the swing equations allow us to determine
the local rates $\omega_k=\dot\theta_k$ of phase angle changes. %, which we
We refer to these deviations
simply as ``local frequencies'' in the following. Their distributions are analyzed in this section.

To identify possible differences for periods of
    weak and strong wind, we perform our analysis conditioned on the
    average wind speed in one-minute intervals. Thus we introduce bins, where the wind speed averaged over one 
    minute lies in ranges 4-6~m/s, 6-8~m/s, $\ldots$, 18-20~m/s. 
    A section of the wind speed series $v(t)$ is shown in Fig.~\ref{fig:windparts}, where we have also indicated
    the division into one-minute segments (each having its own one-minute average wind speed). The wind 
    speed series $v(t)$ is transformed into a power series $P_j^{\rm w}(t)$ as described above.
    
Within the one-minute intervals,
    the wind speed shows strong fluctuations, including sudden changes reflecting its intermittent nature.
    This is demonstrated in Fig.~\ref{fig:increments}(a), where we show the distributions $\psi(\Delta v)$ of one-second wind speed
increments $\Delta v=v(t+1)-v(t)$ obtained from a sampling restricted to
all one-minute sections with
average wind speeds in the bin 8-10~m/s.
These distributions show the typical
heavy tails reflecting the intermittent behavior: large
wind speed changes occur much more frequently than expected from a Gaussian
distribution [green line in Fig.~\ref{fig:increments}(a)]. The heavy tails in the
distribution of wind speed increments are rather directly transformed into heavy tails
of the distribution of wind power increments, see Fig.~\ref{fig:increments}(b).

%To identify possible differences for periods of
%weak and strong wind, we perform our analysis conditioned on the
%average wind speed. Thus we introduce bins, where the one-minute-average wind speed
%lies in ranges 4-6~m/s, 6-8~m/s, $\ldots$, 18-20~m/s. Examples of one-minute segments
%with different mean speeds are shown in Fig.~\ref{fig:windparts}.

Figure~\ref{fig:freqhist}(a) displays an example of a histogram of frequencies, which was
obtained for one realization of one-minute wind power feed-in at node 13 with a mean wind speed belonging to the
bin 8-10~m/s [corresponding to the distribution of wind power increments shown
in Fig.~\ref{fig:increments}(b)]. For this histogram, the local frequencies at all other nodes were sampled.
The shape of the histogram can be described by a
Gaussian core part for small frequencies (solid green line) and a nearly exponential tail behavior for
large frequencies (for both positive and negative deviations from the nominal frequency).
It is interesting to compare this histogram with the one for the homogenized grid structure
when exactly the same sequence of one-minute wind power data is injected at node 13.
The corresponding histogram is shown in Fig.~\ref{fig:freqhist}(b).
Its shape can be described in the same way, but the Gaussian core part has a much smaller width and the
exponential tails decay more rapidly.

When investigating other one-minute sets of wind power data and/or other injection nodes
we obtain histograms of similar shape, which can be characterized
by a Gaussian core and exponential tail part. This holds true irrespective of the
scale of average wind speed, i.e.\ irrespective of the bin, to which the set of one-minute
wind power data is assigned. That similar histogram shapes are obtained for all one-minute sets
is somewhat surprising
in view of the intermittent wind speed behavior, which is reflected in occasional jump-like
changes of the wind speed in short time intervals.\cite{} For example, consider the
one-minute set of wind speed data between 120~s and 180~s and the following set between
180~s and 240~s in Fig.~\ref{fig:windparts}. In the former set, the wind speeds show
only small fluctuations around the mean wind speed. In the latter set, by contrast, jump-like changes
are seen at its beginning and end, and between these sudden changes
there is a strong overall drift from smaller wind speeds of order 6~m/s to larger values of about
10~m/s. The differences in the behaviors of the wind speed thus do not translate into
distinct types of histogram shapes but a different significance of the Gaussian core
and exponential tail part.

We quantify the differences by
introducing two parameters. The first is the standard deviation $\sigma$ of the local
frequency fluctuations, whose value is largely determined by the Gaussian core part.
In fact, the solid green lines in Figs.~\ref{fig:freqhist}(a), (b) correspond to a Gaussian distribution
with zero mean and standard deviation $\sigma$. The second parameter is
\begin{equation}
\alpha=\frac{\langle\omega^4\rangle}{3\sigma^4}-1\,,
\label{eq:alpha}
\end{equation}
which is commonly referred to as the non-Gaussian parameter
in the literature. For a Gaussian distribution one finds $\alpha=0$, while
the exponential tails lead to $\alpha>0$, i.e.\ this parameter quantifies the significance of the
non-Gaussian tails.

For the histogram in Fig.~\ref{fig:freqhist}(a), we find $\sigma=0.13$ and $\alpha=2.83$,
while for the corresponding histogram of the homogenized grid in Fig.~\ref{fig:freqhist}(b)
$\sigma=0.06$ and $\alpha=0.36$. The intermittent nature of the wind must be reflected
in variations of $\sigma$ and $\alpha$ for different one-minute data sets.
These variations are exemplified in the graphs right to the histograms shown in
Fig.~\ref{fig:freqhist}(a) and (b) for 100 sets
belonging to the same bin 8-10~m/s of average wind speed.
The mean values  $\bar\sigma_j$ and $\bar\alpha_j$ are indicated by the horizontal
green lines in these graphs.

\begin{figure}[b]
    \includegraphics[width=0.35\textwidth]{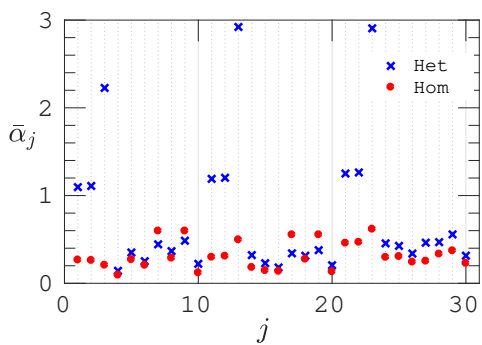}
    \caption{Mean non-Gaussian parameter $\bar\alpha_j$ for the different wind injection nodes $j=1,\ldots,30$
        in the bin of mean wind velocities 8-10~m/s. For each injection node, the averaging was performed over the same
        100 one-minute sets of wind speed data. Blue crosses mark the results for the heterogeneous grid structure and red circles for the homogenized variant.\label{fig:alpha}}
\end{figure}

\begin{figure}[b]
\includegraphics[width=0.48\textwidth]{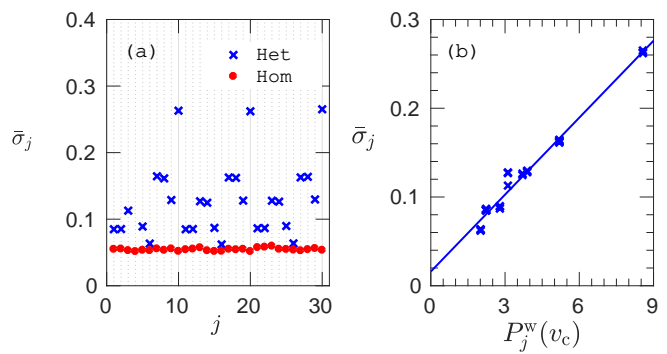}
\caption{(a) Mean values $\bar\sigma_j$ of the standard deviation for the different wind injection nodes $j=1,\ldots,30$ in the bin of mean wind velocities 8-10~m/s. Blue crosses mark the results for the heterogeneous grid structure and red circles for the homogenized variant. (b) Linear relation between $\bar\sigma_j$ and $P_j^{\rm w}(v_c)$
[cf.\ Eq.~\eqref{eq:powerfeedin}] with $v_{\rm c}=9$~m/s the wind speed in the bin center.
\label{fig:sigma}}
\end{figure}

So far we have considered just one injection node $j=13$. To investigate how the
$\bar\sigma$ and $\bar\alpha$ vary with the location of wind power feed-in,
we have determined, for the same hundred sets taken for Figs.~\ref{fig:freqhist}(a) and (b),
the $\bar\sigma_j$ and $\bar\alpha_j$ for all replacements of conventional
generator nodes $j=1,\ldots,30$ by a wind farm.
The results %, obtained by averaging over the same 100 data sets as in
%Figs.~\ref{fig:freqhist}, 
are shown in Figs.~\ref{fig:alpha} and \ref{fig:sigma}.

In Fig.~\ref{fig:alpha} we see that there are two injection nodes $j=13$ and 23 in the heterogeneous
grid (blue crosses) with a large $\alpha_j\simeq3$.
When looking for peculiarities of these nodes, we find that they are the only dead ends
in the IEEE RTS-96, see Fig.~\ref{fig:model-illustration}. In the homogenized
grid this effect of dead ends is not significant and all $\alpha_j\lesssim0.6$
(red circles in Fig.~\ref{fig:alpha}).

For most of the other injection nodes $j$ in the heterogeneous grid we also find
$\alpha_j\lesssim0.6$, with some further exceptions: Node $j=3$ with
$\alpha_j\simeq2.2$ is effectively
a dead end with one strong link to node 38 and a comparatively much weaker link to node 47.
The nodes 1, 2, 11, 12, 21, and 22 with $\alpha_j\gtrsim1$ all belong to
strongly linked pairs (1,2), (11,12), and (21, 22) that are only weakly linked to other nodes. For other bins than the 8-10~m/s bin discussed here, the same nodes $ j $ are identified as those with particular large $ \bar{\alpha}_{j} $ values. Large
non-Gaussian parameters thus occur for injection nodes that are weakly
connected to the entire grid structure.

Interestingly, dead ends were reported as a potentially
destabilizing factor of power grids also when analyzing the attraction basin of fixed points of
Eqs.~\eqref{eq:swing} under frequency and voltage angle perturbations.\cite{Menck/etal:2014, Wolff/etal:2018}
Moreover, the pattern of the $\bar\alpha_j$ seen in Fig.~\ref{fig:alpha} for the heterogeneous case
correlates strongly with an estimation of probabilities of wind injection nodes to give rise to
transmission line overloads.\cite{Schiel/etal:2017} This a remarkable finding, as the estimation of these
overload probabilities was based on a quasi-stationary approach, i.e.\ without considering the
complex nonlinear dynamics given by Eqs.~\eqref{eq:swing}.

The data for the standard deviation $\bar\sigma_j$ in Fig.~\ref{fig:sigma}(a) show again that
the homogenized variant misses to capture important features of the dynamics
seen in the heterogeneous grid structure. While $\bar\sigma_j\simeq0.07$ for all nodes
in the homogenized variant, the $\bar\sigma_j$ in the heterogeneous grid are always larger and
vary significantly between different injection nodes, attaining up to four times larger values.
These variations of the $\bar\sigma_j$ are not related to peculiar topological features, but are simply
connected to the mean wind power injected at node $j$, i.e.\ to the size of the
wind farm, see the discussion in Sec.~\ref{subsec:modeling-stochastic-generation}.
The relation is demonstrated in Fig.~\ref{fig:sigma}(b), where we plotted the $\bar\sigma_j$
against $P_j^{\rm w}(v_c)$ from Eq.~\eqref{eq:powerfeedin}, with $v_{\rm c}=9$~m/s
the wind speed in the center of the considered bin. The linear relation between $\bar\sigma_j$
and $P_j^{\rm w}(v_c)$ is valid also for the other bins of average wind speed.

\section{SUMMARY, CONCLUSIONS AND OUTLOOK}
\label{sec:conclusions}
In summary we have developed an approach to access the impact of short-term
fluctuations of wind speed and associated wind power on local frequency fluctuations in power grids.
For our analysis we used the swing equations by applying a synchronous machine model for all nodes.
We accounted for external control measures by adjusting the grid state to the
average power feed-in on a time scale of one minute. Wind speeds were translated into wind
power by resorting to the so-called power-curve. We considered single-node injection of wind power
at different locations in the IEEE RTS-96 by replacing one of its conventional
generator nodes by a wind farm. The wind farm size was adjusted to the
power of the replaced conventional generator. Modeling of the
wind speeds was based on measured data at a tower in the North Sea with a time resolution of
one second and a stochastic interpolation to smaller time steps used for integrating the swing equations.
In addition we studied the consequences of homogenizing transmission line and node properties.

We found that histograms of local frequency deviations
from the nominal frequency of the synchronous state
exhibit a common type of shape, which we could
describe by a Gaussian distribution for small frequency deviations
and an approximately exponentially decaying tail part.
Without attempting to fit the shape of each individual of the
about 10$^4$ histograms investigated, we 
quantified the relevant features by introducing just two parameters:
(i) the standard deviation of the local frequency fluctuation, which
is largely determined by the Gaussian core part,
and (ii) the non-Gaussian parameter [Eq.~\eqref{eq:alpha}], which
is sensitive to the tail part. The non-Gaussian parameter can assume large values
in the heterogeneous grid for injection nodes
weakly connected to the entire grid, in particular those forming dead ends.
This indicates that these injection nodes are more likely to cause large
local frequency fluctuations. These large fluctuations should be avoided as they can nucleate 
severe grid failures. In a homogenized grid variant, the ``dead end effect''
is not significant. The standard deviations showed strong variations
from node to node. They turned out to be essentially proportional
to the magnitude of the average wind power injection. In agreement with this observation,
the standard deviations were almost the same for all
wind power injection nodes in the homogenized grid variant.

We consider our study to be just a first step for a better understanding of the impact of
short-term fluctuations of wind energy on the stability of power grids. An important task
is to investigate, to which extent different possible origins contribute to
the non-Gaussian features, and whether one can identify a dominating one.
To this end it will be useful to separate effects associated with (i) the nonlinearities
in the swing equations [Eq.~\eqref{eq:swing}] , (ii) the saturation of the power curve
[Eq.~\eqref{eq:powerfeedin}], and (iii) the non-Gaussian wind statistics, in particular its
intermittent nature implied by atmospheric turbulence, as reflected in the increment
distributions shown in Figs.~\ref{fig:increments}(a) and \ref{fig:increments}(b).
 In this context, improvements
in the modeling should be implemented for a more realistic stochastic translation of wind speed into power
\cite{Milan/etal:2014} and a better modeling of the wind injection, e.g.\  by using measured
data with finer resolution (if available), or by
applying reliable models for generating surrogate data, or by improving the stochastic
interpolation method between measured data.
Further developments should include also a more realistic representation of the load nodes, e.g.\ by using the
effective network or structure preserving model,\cite{Nishikawa/Motter:2015} as well as an account for
the modified dynamics implied by the ac/dc and dc/ac inverters used for the wind power feed-in.
The swing equations~\eqref{eq:swing} can be modified to take into account  secondary control measures
also,\cite{Boettcher/etal:2019} which would allow one to extend the analysis to longer time scales.

Going along with these improvements in the grid modeling, extended setups of the wind power feed-in
need to be studied and the analysis of stochastic grid dynamics should be widened. As for
the wind power feed-in, higher penetrations of the grid with wind power must be investigated
under consideration of both spatial and temporal correlations of wind velocities.
As for the analysis of stochastic grid dynamics, it is important to quantify
correlations between local frequency fluctuations. Preliminary results for distributions of frequency
increments show similar features as reported for the histograms of the
frequencies themselves.

\begin{acknowledgments}
Financial support from the Deutsche Forschungsgemeinschaft (MA 1636/9-1 and PE 478/16-1) is gratefully acknowledged.
\end{acknowledgments}

\appendix

\section{Stochastic Interpolation}
Let $v_0$ and $v_1$ be two of the measured wind speeds separated by one second.
Defining two stochastic processes $v_+(t)$ and $v_-(t)$, $t\in[0,1]$,  with starting  values $v_+(0)=v_0$, $v_-(0)=v_1$, a stochastic interpolation between $v_0$ and $v_1$ is given by
\begin{equation}
v(t)=(1-t)v_+(t)+tv_-(1-t)\,,\quad t\in[0,1]\,.
\label{eq:v-vpvm}
\end{equation}
Hence the process $v_+(t)$ is considered to run forward in time and to contribute to $v(t)$ with weight $(1-t)$,
while the process $v_-(t)$ runs backward in time and contributes with weight $t$.

Specifically, we use here a simple Ornstein-Uhlenbeck
process for $v_\pm(t)$ , i.e.\ $v_\pm(t)$ obey the Langevin equations
\begin{equation}
\frac{\mathrm{d} v_\pm(t)}{\mathrm{d}t}=-\gamma[v_\pm(t)-\bar v_\pm(t)]+\sqrt{2\Gamma}\,\eta(t)\,,
\label{eq:vpm-processes}
\end{equation}
where $\eta(t)$ is a Gaussian white noise with zero mean and correlator $\langle\eta(t)\eta(t')\rangle=\delta(t-t')$.
Based on studies of correlation properties of the measured time series $\{v_n\}_{1\le n\le N}$,
\footnote{M.\ Brune, ``Modeling of wind speeds with the Langevin equation'' (in German), bachelor thesis (2018), Osnabr{\"u}ck University.}
%\cite{Brune:2018}
we set $\gamma$ to $0.54$ and $\Gamma=\gamma/2$.
The time-dependent mean values $\bar v_\pm(t)$ are $\bar v_+(t)=\bar v_-(1-t)=\bar v(t)$,
where
\begin{equation}
\bar v(t)=v_0+(v_1-v_0)t\,,\quad t\in[0,1]\,,
\end{equation}
is the linear interpolation
between $v_0$ and $v_1$. Hence, the Ornstein-Uhlenbeck processes $v_+(t)$ and $v_+(1-t)$ in
Eq.~\eqref{eq:v-vpvm} are biased towards the
linearly interpolated value at all intermediate times $t$.

The Langevin equations \eqref{eq:vpm-processes}
are integrated to yield a stochastic sequence $v(n\Delta t)$, $n=1,\ldots,(1/\Delta t-1)$, of wind speeds between
$v_0$ and $v_1$ with time resolution $\Delta t$, where in our case $\Delta t=5\times 10^{-4}$s.

%\bibliographystyle{apsrev4-1}
%\bibliography{power-grid-min}% Produces the bibliography via BibTeX.

\begin{thebibliography}{41}%
\makeatletter
\providecommand \@ifxundefined [1]{%
 \@ifx{#1\undefined}
}%
\providecommand \@ifnum [1]{%
 \ifnum #1\expandafter \@firstoftwo
 \else \expandafter \@secondoftwo
 \fi
}%
\providecommand \@ifx [1]{%
 \ifx #1\expandafter \@firstoftwo
 \else \expandafter \@secondoftwo
 \fi
}%
\providecommand \natexlab [1]{#1}%
\providecommand \enquote  [1]{``#1''}%
\providecommand \bibnamefont  [1]{#1}%
\providecommand \bibfnamefont [1]{#1}%
\providecommand \citenamefont [1]{#1}%
\providecommand \href@noop [0]{\@secondoftwo}%
\providecommand \href [0]{\begingroup \@sanitize@url \@href}%
\providecommand \@href[1]{\@@startlink{#1}\@@href}%
\providecommand \@@href[1]{\endgroup#1\@@endlink}%
\providecommand \@sanitize@url [0]{\catcode `\\12\catcode `\$12\catcode
  `\&12\catcode `\#12\catcode `\^12\catcode `\_12\catcode `\%12\relax}%
\providecommand \@@startlink[1]{}%
\providecommand \@@endlink[0]{}%
\providecommand \url  [0]{\begingroup\@sanitize@url \@url }%
\providecommand \@url [1]{\endgroup\@href {#1}{\urlprefix }}%
\providecommand \urlprefix  [0]{URL }%
\providecommand \Eprint [0]{\href }%
\providecommand \doibase [0]{http://dx.doi.org/}%
\providecommand \selectlanguage [0]{\@gobble}%
\providecommand \bibinfo  [0]{\@secondoftwo}%
\providecommand \bibfield  [0]{\@secondoftwo}%
\providecommand \translation [1]{[#1]}%
\providecommand \BibitemOpen [0]{}%
\providecommand \bibitemStop [0]{}%
\providecommand \bibitemNoStop [0]{.\EOS\space}%
\providecommand \EOS [0]{\spacefactor3000\relax}%
\providecommand \BibitemShut  [1]{\csname bibitem#1\endcsname}%
\let\auto@bib@innerbib\@empty
%</preamble>
\bibitem [{\citenamefont {Menck}\ \emph {et~al.}(2014)\citenamefont {Menck},
  \citenamefont {Heitzig}, \citenamefont {Kurths},\ and\ \citenamefont
  {Schellnhuber}}]{Menck/etal:2014}%
  \BibitemOpen
  \bibfield  {author} {\bibinfo {author} {\bibfnamefont {P.}~\bibnamefont
  {Menck}}, \bibinfo {author} {\bibfnamefont {J.}~\bibnamefont {Heitzig}},
  \bibinfo {author} {\bibfnamefont {J.}~\bibnamefont {Kurths}}, \ and\ \bibinfo
  {author} {\bibfnamefont {H.}~\bibnamefont {Schellnhuber}},\ }\href@noop {}
  {\bibfield  {journal} {\bibinfo  {journal} {Nat. Commun.}\ }\textbf {\bibinfo
  {volume} {5}},\ \bibinfo {pages} {3969} (\bibinfo {year} {2014})}\BibitemShut
  {NoStop}%
\bibitem [{Sch()}]{Schultz/etal:2014}%
  \BibitemOpen
  \href@noop {} {}\bibinfo {note} {P.\ Schultz, J.\ Heitzig, and J.\ Kurths,
  Eur.\ Phys.\ J.: Spec.\ Top.\ \textbf{225}, 2593 (2014)}\BibitemShut
  {NoStop}%
\bibitem [{\citenamefont {Jung}\ and\ \citenamefont
  {Kettemann}(2016)}]{Jung/Kettemann:2016}%
  \BibitemOpen
  \bibfield  {author} {\bibinfo {author} {\bibfnamefont {D.}~\bibnamefont
  {Jung}}\ and\ \bibinfo {author} {\bibfnamefont {S.}~\bibnamefont
  {Kettemann}},\ }\href@noop {} {\bibfield  {journal} {\bibinfo  {journal}
  {Phys. Rev. E}\ }\textbf {\bibinfo {volume} {94}},\ \bibinfo {pages} {012307}
  (\bibinfo {year} {2016})}\BibitemShut {NoStop}%
\bibitem [{\citenamefont {Auer}\ \emph {et~al.}(2016)\citenamefont {Auer},
  \citenamefont {Kleis}, \citenamefont {Schultz}, \citenamefont {Kurths},\ and\
  \citenamefont {Hellmann}}]{Auer/etal:2016}%
  \BibitemOpen
  \bibfield  {author} {\bibinfo {author} {\bibfnamefont {S.}~\bibnamefont
  {Auer}}, \bibinfo {author} {\bibfnamefont {K.}~\bibnamefont {Kleis}},
  \bibinfo {author} {\bibfnamefont {P.}~\bibnamefont {Schultz}}, \bibinfo
  {author} {\bibfnamefont {J.}~\bibnamefont {Kurths}}, \ and\ \bibinfo {author}
  {\bibfnamefont {F.}~\bibnamefont {Hellmann}},\ }\href {\doibase
  10.1140/epjst/e2015-50265-9} {\bibfield  {journal} {\bibinfo  {journal} {Eur.
  Phys. J. Spec. Top.}\ }\textbf {\bibinfo {volume} {225}},\ \bibinfo {pages}
  {609} (\bibinfo {year} {2016})}\BibitemShut {NoStop}%
\bibitem [{\citenamefont {Auer}\ \emph {et~al.}(2017)\citenamefont {Auer},
  \citenamefont {Hellmann}, \citenamefont {Krause},\ and\ \citenamefont
  {Kurths}}]{Auer/etal:2017}%
  \BibitemOpen
  \bibfield  {author} {\bibinfo {author} {\bibfnamefont {S.}~\bibnamefont
  {Auer}}, \bibinfo {author} {\bibfnamefont {F.}~\bibnamefont {Hellmann}},
  \bibinfo {author} {\bibfnamefont {M.}~\bibnamefont {Krause}}, \ and\ \bibinfo
  {author} {\bibfnamefont {J.}~\bibnamefont {Kurths}},\ }\href {\doibase
  10.1063/1.5001818} {\bibfield  {journal} {\bibinfo  {journal} {Chaos}\
  }\textbf {\bibinfo {volume} {27}},\ \bibinfo {pages} {127003} (\bibinfo
  {year} {2017})},\ \Eprint
  {http://arxiv.org/abs/https://doi.org/10.1063/1.5001818}
  {https://doi.org/10.1063/1.5001818} \BibitemShut {NoStop}%
\bibitem [{\citenamefont {Schiel}\ \emph {et~al.}(2017)\citenamefont {Schiel},
  \citenamefont {Lind},\ and\ \citenamefont {Maass}}]{Schiel/etal:2017}%
  \BibitemOpen
  \bibfield  {author} {\bibinfo {author} {\bibfnamefont {C.}~\bibnamefont
  {Schiel}}, \bibinfo {author} {\bibfnamefont {P.~G.}\ \bibnamefont {Lind}}, \
  and\ \bibinfo {author} {\bibfnamefont {P.}~\bibnamefont {Maass}},\ }\href
  {\doibase 10.1038/s41598-017-11465-w} {\bibfield  {journal} {\bibinfo
  {journal} {Sci. Rep.}\ }\textbf {\bibinfo {volume} {7}},\ \bibinfo {pages}
  {11562} (\bibinfo {year} {2017})}\BibitemShut {NoStop}%
\bibitem [{\citenamefont {Wolff}\ \emph {et~al.}(2018)\citenamefont {Wolff},
  \citenamefont {Lind},\ and\ \citenamefont {Maass}}]{Wolff/etal:2018}%
  \BibitemOpen
  \bibfield  {author} {\bibinfo {author} {\bibfnamefont {M.~F.}\ \bibnamefont
  {Wolff}}, \bibinfo {author} {\bibfnamefont {P.~G.}\ \bibnamefont {Lind}}, \
  and\ \bibinfo {author} {\bibfnamefont {P.}~\bibnamefont {Maass}},\ }\href
  {\doibase 10.1063/1.5040689} {\bibfield  {journal} {\bibinfo  {journal}
  {Chaos}\ }\textbf {\bibinfo {volume} {28}},\ \bibinfo {pages} {103120}
  (\bibinfo {year} {2018})}\BibitemShut {NoStop}%
\bibitem [{\citenamefont {Schultz}\ \emph {et~al.}(2014)\citenamefont
  {Schultz}, \citenamefont {Heitzig},\ and\ \citenamefont
  {Kurths}}]{Schultz/etal:2014a}%
  \BibitemOpen
  \bibfield  {author} {\bibinfo {author} {\bibfnamefont {P.}~\bibnamefont
  {Schultz}}, \bibinfo {author} {\bibfnamefont {J.}~\bibnamefont {Heitzig}}, \
  and\ \bibinfo {author} {\bibfnamefont {J.}~\bibnamefont {Kurths}},\
  }\href@noop {} {\bibfield  {journal} {\bibinfo  {journal} {New J. Phys.}\
  }\textbf {\bibinfo {volume} {16}},\ \bibinfo {pages} {125001} (\bibinfo
  {year} {2014})}\BibitemShut {NoStop}%
\bibitem [{\citenamefont {Kim}\ \emph {et~al.}(2018)\citenamefont {Kim},
  \citenamefont {Lee}, \citenamefont {Davidsen},\ and\ \citenamefont
  {Son}}]{Kim/etal:2018}%
  \BibitemOpen
  \bibfield  {author} {\bibinfo {author} {\bibfnamefont {H.}~\bibnamefont
  {Kim}}, \bibinfo {author} {\bibfnamefont {S.~H.}\ \bibnamefont {Lee}},
  \bibinfo {author} {\bibfnamefont {J.}~\bibnamefont {Davidsen}}, \ and\
  \bibinfo {author} {\bibfnamefont {S.-W.}\ \bibnamefont {Son}},\ }\href@noop
  {} {\bibfield  {journal} {\bibinfo  {journal} {New J. Phys.}\ }\textbf
  {\bibinfo {volume} {20}},\ \bibinfo {pages} {113006} (\bibinfo {year}
  {2018})}\BibitemShut {NoStop}%
\bibitem [{\citenamefont {M{\"u}ller}\ \emph {et~al.}(2018)\citenamefont
  {M{\"u}ller}, \citenamefont {Wienholt}, \citenamefont {Kleinhans},
  \citenamefont {Cussmann}, \citenamefont {Bunke}, \citenamefont
  {Ple{\ss}mann},\ and\ \citenamefont {Wendiggensen}}]{Mueller/etal:2018}%
  \BibitemOpen
  \bibfield  {author} {\bibinfo {author} {\bibfnamefont {U.~P.}\ \bibnamefont
  {M{\"u}ller}}, \bibinfo {author} {\bibfnamefont {L.}~\bibnamefont
  {Wienholt}}, \bibinfo {author} {\bibfnamefont {D.}~\bibnamefont {Kleinhans}},
  \bibinfo {author} {\bibfnamefont {I.}~\bibnamefont {Cussmann}}, \bibinfo
  {author} {\bibfnamefont {W.-D.}\ \bibnamefont {Bunke}}, \bibinfo {author}
  {\bibfnamefont {G.}~\bibnamefont {Ple{\ss}mann}}, \ and\ \bibinfo {author}
  {\bibfnamefont {J.}~\bibnamefont {Wendiggensen}},\ }\href {\doibase
  10.1088/1742-6596/977/1/012003} {\bibfield  {journal} {\bibinfo  {journal}
  {J. Phys.: Conf. Ser.}\ }\textbf {\bibinfo {volume} {977}},\ \bibinfo {pages}
  {012003} (\bibinfo {year} {2018})}\BibitemShut {NoStop}%
\bibitem [{Sci()}]{SciGRIDv0.2}%
  \BibitemOpen
  \href {http://www.scigrid.de} {}\bibinfo {note} {C. Matke, W. Medjroubi, and
  D. Kleinhans, {SciGRID} - {A}n {O}pen {S}ource {R}eference {M}odel for the
  {E}uropean {T}ransmission {N}etwork (v0.2), 2016.}\BibitemShut {Stop}%
\bibitem [{Ang()}]{Anghel/etal:2007}%
  \BibitemOpen
  \href@noop {} {}\bibinfo {note} {M.\ Anghel, K.\ A.\ Werley, and A.\ E.\
  Motter, \textit{40th Annual Hawaii International Conference on System
  Sciences (HICSS'07)} (IEEE Computer Society, 2007).
  pp.~2174-2180}\BibitemShut {NoStop}%
\bibitem [{\citenamefont {Chertkov}\ \emph {et~al.}(2011)\citenamefont
  {Chertkov}, \citenamefont {Stepanov}, \citenamefont {Pan},\ and\
  \citenamefont {Baldick}}]{Chertkov/etal:2011}%
  \BibitemOpen
  \bibfield  {author} {\bibinfo {author} {\bibfnamefont {M.}~\bibnamefont
  {Chertkov}}, \bibinfo {author} {\bibfnamefont {M.}~\bibnamefont {Stepanov}},
  \bibinfo {author} {\bibfnamefont {F.}~\bibnamefont {Pan}}, \ and\ \bibinfo
  {author} {\bibfnamefont {R.}~\bibnamefont {Baldick}},\ }in\ \href@noop {}
  {\emph {\bibinfo {booktitle} {Proceedings of the 50th IEEE Conference on
  Decision and Control and European Control Conference, CDC-ECC 2011 - Orlando,
  FL, United States}}}\ (\bibinfo {year} {2011})\ pp.\ \bibinfo {pages}
  {2174--2180}\BibitemShut {NoStop}%
\bibitem [{\citenamefont {Schmietendorf}\ \emph {et~al.}(2014)\citenamefont
  {Schmietendorf}, \citenamefont {Peinke}, \citenamefont {Friedrich},\ and\
  \citenamefont {Kamps}}]{Schmietendorf/etal:2014}%
  \BibitemOpen
  \bibfield  {author} {\bibinfo {author} {\bibfnamefont {K.}~\bibnamefont
  {Schmietendorf}}, \bibinfo {author} {\bibfnamefont {J.}~\bibnamefont
  {Peinke}}, \bibinfo {author} {\bibfnamefont {R.}~\bibnamefont {Friedrich}}, \
  and\ \bibinfo {author} {\bibfnamefont {O.}~\bibnamefont {Kamps}},\
  }\href@noop {} {\bibfield  {journal} {\bibinfo  {journal} {Eur. Phys. J.
  Spec. Top.}\ }\textbf {\bibinfo {volume} {223}},\ \bibinfo {pages} {2577}
  (\bibinfo {year} {2014})}\BibitemShut {NoStop}%
\bibitem [{\citenamefont {Nishikawa}\ and\ \citenamefont
  {Motter}(2015)}]{Nishikawa/Motter:2015}%
  \BibitemOpen
  \bibfield  {author} {\bibinfo {author} {\bibfnamefont {T.}~\bibnamefont
  {Nishikawa}}\ and\ \bibinfo {author} {\bibfnamefont {A.}~\bibnamefont
  {Motter}},\ }\href@noop {} {\bibfield  {journal} {\bibinfo  {journal} {New J.
  Phys.}\ }\textbf {\bibinfo {volume} {17}},\ \bibinfo {pages} {015012}
  (\bibinfo {year} {2015})}\BibitemShut {NoStop}%
\bibitem [{\citenamefont {Schmietendorf}\ \emph {et~al.}(2017)\citenamefont
  {Schmietendorf}, \citenamefont {Peinke},\ and\ \citenamefont
  {Kamps}}]{Schmietendorf/etal:2017}%
  \BibitemOpen
  \bibfield  {author} {\bibinfo {author} {\bibfnamefont {K.}~\bibnamefont
  {Schmietendorf}}, \bibinfo {author} {\bibfnamefont {J.}~\bibnamefont
  {Peinke}}, \ and\ \bibinfo {author} {\bibfnamefont {O.}~\bibnamefont
  {Kamps}},\ }\href {\doibase 10.1140/epjb/e2017-80352-8} {\bibfield  {journal}
  {\bibinfo  {journal} {Eur. Phys. J. B}\ }\textbf {\bibinfo {volume} {90}},\
  \bibinfo {pages} {222} (\bibinfo {year} {2017})}\BibitemShut {NoStop}%
\bibitem [{\citenamefont {H{\"a}hne}\ \emph {et~al.}(2018)\citenamefont
  {H{\"a}hne}, \citenamefont {Schottler}, \citenamefont {W{\"a}chter},
  \citenamefont {Peinke},\ and\ \citenamefont {Kamps}}]{Haehne/etal:2018}%
  \BibitemOpen
  \bibfield  {author} {\bibinfo {author} {\bibfnamefont {H.}~\bibnamefont
  {H{\"a}hne}}, \bibinfo {author} {\bibfnamefont {J.}~\bibnamefont
  {Schottler}}, \bibinfo {author} {\bibfnamefont {M.}~\bibnamefont
  {W{\"a}chter}}, \bibinfo {author} {\bibfnamefont {J.}~\bibnamefont {Peinke}},
  \ and\ \bibinfo {author} {\bibfnamefont {O.}~\bibnamefont {Kamps}},\ }\href
  {http://stacks.iop.org/0295-5075/121/i=3/a=30001} {\bibfield  {journal}
  {\bibinfo  {journal} {EPL}\ }\textbf {\bibinfo {volume} {121}},\ \bibinfo
  {pages} {30001} (\bibinfo {year} {2018})}\BibitemShut {NoStop}%
\bibitem [{\citenamefont {H\"ahne}\ \emph {et~al.}(2019)\citenamefont
  {H\"ahne}, \citenamefont {Schmietendorf}, \citenamefont {Tamrakar},
  \citenamefont {Peinke},\ and\ \citenamefont {Kettemann}}]{Haehne/etal:2019}%
  \BibitemOpen
  \bibfield  {author} {\bibinfo {author} {\bibfnamefont {H.}~\bibnamefont
  {H\"ahne}}, \bibinfo {author} {\bibfnamefont {K.}~\bibnamefont
  {Schmietendorf}}, \bibinfo {author} {\bibfnamefont {S.}~\bibnamefont
  {Tamrakar}}, \bibinfo {author} {\bibfnamefont {J.}~\bibnamefont {Peinke}}, \
  and\ \bibinfo {author} {\bibfnamefont {S.}~\bibnamefont {Kettemann}},\ }\href
  {\doibase 10.1103/PhysRevE.99.050301} {\bibfield  {journal} {\bibinfo
  {journal} {Phys. Rev. E}\ }\textbf {\bibinfo {volume} {99}},\ \bibinfo
  {pages} {050301} (\bibinfo {year} {2019})}\BibitemShut {NoStop}%
\bibitem [{\citenamefont {Sch{\"a}fer}\ \emph
  {et~al.}(2018{\natexlab{a}})\citenamefont {Sch{\"a}fer}, \citenamefont
  {Beck}, \citenamefont {Aihara}, \citenamefont {Witthaut},\ and\ \citenamefont
  {Timme}}]{Schaefer/etal:2018}%
  \BibitemOpen
  \bibfield  {author} {\bibinfo {author} {\bibfnamefont {B.}~\bibnamefont
  {Sch{\"a}fer}}, \bibinfo {author} {\bibfnamefont {C.}~\bibnamefont {Beck}},
  \bibinfo {author} {\bibfnamefont {K.}~\bibnamefont {Aihara}}, \bibinfo
  {author} {\bibfnamefont {D.}~\bibnamefont {Witthaut}}, \ and\ \bibinfo
  {author} {\bibfnamefont {M.}~\bibnamefont {Timme}},\ }\href {\doibase
  10.1038/s41560-017-0058-z} {\bibfield  {journal} {\bibinfo  {journal} {Nat.
  Energy}\ }\textbf {\bibinfo {volume} {3}},\ \bibinfo {pages} {119} (\bibinfo
  {year} {2018}{\natexlab{a}})}\BibitemShut {NoStop}%
\bibitem [{\citenamefont {Albert}\ \emph {et~al.}(2004)\citenamefont {Albert},
  \citenamefont {Albert},\ and\ \citenamefont {Nakarado}}]{Albert/etal:2004}%
  \BibitemOpen
  \bibfield  {author} {\bibinfo {author} {\bibfnamefont {R.}~\bibnamefont
  {Albert}}, \bibinfo {author} {\bibfnamefont {I.}~\bibnamefont {Albert}}, \
  and\ \bibinfo {author} {\bibfnamefont {G.~L.}\ \bibnamefont {Nakarado}},\
  }\href {\doibase 10.1103/PhysRevE.69.025103} {\bibfield  {journal} {\bibinfo
  {journal} {Phys. Rev. E}\ }\textbf {\bibinfo {volume} {69}},\ \bibinfo
  {pages} {025103} (\bibinfo {year} {2004})}\BibitemShut {NoStop}%
\bibitem [{\citenamefont {Wang}\ and\ \citenamefont
  {Rong}(2009)}]{Wang/Rong:2009}%
  \BibitemOpen
  \bibfield  {author} {\bibinfo {author} {\bibfnamefont {J.-W.}\ \bibnamefont
  {Wang}}\ and\ \bibinfo {author} {\bibfnamefont {L.-L.}\ \bibnamefont
  {Rong}},\ }\href {\doibase https://doi.org/10.1016/j.ssci.2009.02.002}
  {\bibfield  {journal} {\bibinfo  {journal} {Saf. Sci.}\ }\textbf {\bibinfo
  {volume} {47}},\ \bibinfo {pages} {1332 } (\bibinfo {year}
  {2009})}\BibitemShut {NoStop}%
\bibitem [{\citenamefont {Hines}\ \emph {et~al.}(2010)\citenamefont {Hines},
  \citenamefont {Cotilla-Sanchez},\ and\ \citenamefont
  {Blumsack}}]{Hines/etal:2010}%
  \BibitemOpen
  \bibfield  {author} {\bibinfo {author} {\bibfnamefont {P.}~\bibnamefont
  {Hines}}, \bibinfo {author} {\bibfnamefont {E.}~\bibnamefont
  {Cotilla-Sanchez}}, \ and\ \bibinfo {author} {\bibfnamefont {S.}~\bibnamefont
  {Blumsack}},\ }\href {https://doi.org/10.1063/1.3489887} {\bibfield
  {journal} {\bibinfo  {journal} {Chaos}\ }\textbf {\bibinfo {volume} {20}},\
  \bibinfo {pages} {033122} (\bibinfo {year} {2010})}\BibitemShut {NoStop}%
\bibitem [{And()}]{Andrychowicz:2016}%
  \BibitemOpen
  \href@noop {} {}\bibinfo {note} {M. Andrychowicz and B. Olek, {\it Optimal
  structure of the RES in distribution systems}, in 13th International
  Conference on the European Energy Market (2016), pp.\ 1-5.}\BibitemShut
  {Stop}%
\bibitem [{\citenamefont {Schmietendorf}\ \emph {et~al.}(2018)\citenamefont
  {Schmietendorf}, \citenamefont {Kamps}, \citenamefont {Wolff}, \citenamefont
  {Lind}, \citenamefont {Maass},\ and\ \citenamefont
  {Peinke}}]{Schmietendorf/etal:2018}%
  \BibitemOpen
  \bibfield  {author} {\bibinfo {author} {\bibfnamefont {K.}~\bibnamefont
  {Schmietendorf}}, \bibinfo {author} {\bibfnamefont {O.}~\bibnamefont
  {Kamps}}, \bibinfo {author} {\bibfnamefont {M.}~\bibnamefont {Wolff}},
  \bibinfo {author} {\bibfnamefont {P.~G.}\ \bibnamefont {Lind}}, \bibinfo
  {author} {\bibfnamefont {P.}~\bibnamefont {Maass}}, \ and\ \bibinfo {author}
  {\bibfnamefont {J.}~\bibnamefont {Peinke}},\ }\href@noop {} {\enquote
  {\bibinfo {title} {Bridging between load-flow and {K}uramoto-like power grid
  models: A flexible approach to integrating electrical storage units},}\
  }\bibinfo {howpublished} {arXiv:1812.01972} (\bibinfo {year}
  {2018})\BibitemShut {NoStop}%
\bibitem [{\citenamefont {Grigg}\ \emph {et~al.}(1999)\citenamefont {Grigg},
  \citenamefont {Wong}, \citenamefont {Albrecht}, \citenamefont {Allan},
  \citenamefont {Bhavaraju}, \citenamefont {Billinton}, \citenamefont {Chen},
  \citenamefont {Fong}, \citenamefont {Haddad}, \citenamefont {Kuruganty},\
  and\ \citenamefont {{et al.}}}]{Grigg/etal:1999}%
  \BibitemOpen
  \bibfield  {author} {\bibinfo {author} {\bibfnamefont {C.}~\bibnamefont
  {Grigg}}, \bibinfo {author} {\bibfnamefont {P.}~\bibnamefont {Wong}},
  \bibinfo {author} {\bibfnamefont {P.}~\bibnamefont {Albrecht}}, \bibinfo
  {author} {\bibfnamefont {R.}~\bibnamefont {Allan}}, \bibinfo {author}
  {\bibfnamefont {M.}~\bibnamefont {Bhavaraju}}, \bibinfo {author}
  {\bibfnamefont {R.}~\bibnamefont {Billinton}}, \bibinfo {author}
  {\bibfnamefont {Q.}~\bibnamefont {Chen}}, \bibinfo {author} {\bibfnamefont
  {C.}~\bibnamefont {Fong}}, \bibinfo {author} {\bibfnamefont {S.}~\bibnamefont
  {Haddad}}, \bibinfo {author} {\bibfnamefont {S.}~\bibnamefont {Kuruganty}}, \
  and\ \bibinfo {author} {\bibnamefont {{et al.}}},\ }\href@noop {} {\bibfield
  {journal} {\bibinfo  {journal} {IEEE Trans. Pow. Sys.}\ }\textbf {\bibinfo
  {volume} {14}},\ \bibinfo {pages} {1010} (\bibinfo {year}
  {1999})}\BibitemShut {NoStop}%
\bibitem [{Note1()}]{Note1}%
  \BibitemOpen
  \bibinfo {note} {FINO I project and database (2016). The FINO project is
  supported by the German Government through BMWi and PTJ. Available at
  \protect \texttt {http://www.bsh.de}}\BibitemShut {NoStop}%
\bibitem [{\citenamefont {Rohden}\ \emph {et~al.}(2012)\citenamefont {Rohden},
  \citenamefont {Sorge}, \citenamefont {Timme},\ and\ \citenamefont
  {Witthaut}}]{Rohden/etal:2012}%
  \BibitemOpen
  \bibfield  {author} {\bibinfo {author} {\bibfnamefont {M.}~\bibnamefont
  {Rohden}}, \bibinfo {author} {\bibfnamefont {A.}~\bibnamefont {Sorge}},
  \bibinfo {author} {\bibfnamefont {M.}~\bibnamefont {Timme}}, \ and\ \bibinfo
  {author} {\bibfnamefont {D.}~\bibnamefont {Witthaut}},\ }\href {\doibase
  10.1103/PhysRevLett.109.064101} {\bibfield  {journal} {\bibinfo  {journal}
  {Phys. Rev. Lett.}\ }\textbf {\bibinfo {volume} {109}},\ \bibinfo {pages}
  {064101} (\bibinfo {year} {2012})}\BibitemShut {NoStop}%
\bibitem [{\citenamefont {Sch{\"a}fer}\ \emph
  {et~al.}(2018{\natexlab{b}})\citenamefont {Sch{\"a}fer}, \citenamefont
  {Witthaut}, \citenamefont {Timme},\ and\ \citenamefont
  {Latora}}]{Schaefer/etal:2018:2}%
  \BibitemOpen
  \bibfield  {author} {\bibinfo {author} {\bibfnamefont {B.}~\bibnamefont
  {Sch{\"a}fer}}, \bibinfo {author} {\bibfnamefont {D.}~\bibnamefont
  {Witthaut}}, \bibinfo {author} {\bibfnamefont {M.}~\bibnamefont {Timme}}, \
  and\ \bibinfo {author} {\bibfnamefont {V.}~\bibnamefont {Latora}},\ }\href
  {\doibase 10.1038/s41467-018-04287-5} {\bibfield  {journal} {\bibinfo
  {journal} {Nat. Commun.}\ }\textbf {\bibinfo {volume} {9}},\ \bibinfo {pages}
  {1975} (\bibinfo {year} {2018}{\natexlab{b}})}\BibitemShut {NoStop}%
\bibitem [{\citenamefont {Tamrakar}\ \emph {et~al.}(2018)\citenamefont
  {Tamrakar}, \citenamefont {Conrath},\ and\ \citenamefont
  {Kettemann}}]{Tamrakar/etal:2018}%
  \BibitemOpen
  \bibfield  {author} {\bibinfo {author} {\bibfnamefont {S.}~\bibnamefont
  {Tamrakar}}, \bibinfo {author} {\bibfnamefont {M.}~\bibnamefont {Conrath}}, \
  and\ \bibinfo {author} {\bibfnamefont {S.}~\bibnamefont {Kettemann}},\ }\href
  {\doibase 10.1038/s41598-018-24685-5} {\bibfield  {journal} {\bibinfo
  {journal} {Sci. Rep.}\ }\textbf {\bibinfo {volume} {8}},\ \bibinfo {pages}
  {6459} (\bibinfo {year} {2018})}\BibitemShut {NoStop}%
\bibitem [{\citenamefont {Machowski}\ \emph {et~al.}(2008)\citenamefont
  {Machowski}, \citenamefont {Bialek},\ and\ \citenamefont
  {Bumby}}]{Machowski/etal:2008}%
  \BibitemOpen
  \bibfield  {author} {\bibinfo {author} {\bibfnamefont {J.}~\bibnamefont
  {Machowski}}, \bibinfo {author} {\bibfnamefont {J.}~\bibnamefont {Bialek}}, \
  and\ \bibinfo {author} {\bibfnamefont {J.}~\bibnamefont {Bumby}},\
  }\href@noop {} {\emph {\bibinfo {title} {Power System Dynamics}}}\ (\bibinfo
  {publisher} {John Wiley \& Sons, New Jersey},\ \bibinfo {year}
  {2008})\BibitemShut {NoStop}%
\bibitem [{\citenamefont {Filatrella}\ \emph {et~al.}(2008)\citenamefont
  {Filatrella}, \citenamefont {Nielsen},\ and\ \citenamefont
  {Pedersen}}]{Filatrella/etal:2008}%
  \BibitemOpen
  \bibfield  {author} {\bibinfo {author} {\bibfnamefont {G.}~\bibnamefont
  {Filatrella}}, \bibinfo {author} {\bibfnamefont {A.~H.}\ \bibnamefont
  {Nielsen}}, \ and\ \bibinfo {author} {\bibfnamefont {N.~F.}\ \bibnamefont
  {Pedersen}},\ }\href {\doibase 10.1140/epjb/e2008-00098-8} {\bibfield
  {journal} {\bibinfo  {journal} {Eur. Phys. J. B}\ }\textbf {\bibinfo {volume}
  {61}},\ \bibinfo {pages} {485} (\bibinfo {year} {2008})}\BibitemShut
  {NoStop}%
\bibitem [{Note2()}]{Note2}%
  \BibitemOpen
  \bibinfo {note} {See \protect \texttt {https://www.entsoe.eu} for ``UCTE
  Operation Handbook'' (2004).}\BibitemShut {Stop}%
\bibitem [{\citenamefont {Kuramoto}(1975)}]{Kuramoto1975}%
  \BibitemOpen
  \bibfield  {author} {\bibinfo {author} {\bibfnamefont {Y.}~\bibnamefont
  {Kuramoto}},\ }in\ \href {\doibase 10.1007/BFb0013365} {\emph {\bibinfo
  {booktitle} {Mathematical Problems in Theoretical Physics}}},\ \bibinfo
  {series} {Lecture Notes in Physics}, Vol.~\bibinfo {volume} {39},\ \bibinfo
  {editor} {edited by\ \bibinfo {editor} {\bibfnamefont {H.}~\bibnamefont
  {{Araki}}}}\ (\bibinfo  {publisher} {Springer Verlag, Berlin},\ \bibinfo
  {year} {1975})\ pp.\ \bibinfo {pages} {420--422}\BibitemShut {NoStop}%
\bibitem [{\citenamefont {Acebr\'on}\ \emph {et~al.}(2005)\citenamefont
  {Acebr\'on}, \citenamefont {Bonilla}, \citenamefont {P\'erez~Vicente},
  \citenamefont {Ritort},\ and\ \citenamefont {Spigler}}]{Acebron/etal:2005}%
  \BibitemOpen
  \bibfield  {author} {\bibinfo {author} {\bibfnamefont {J.~A.}\ \bibnamefont
  {Acebr\'on}}, \bibinfo {author} {\bibfnamefont {L.~L.}\ \bibnamefont
  {Bonilla}}, \bibinfo {author} {\bibfnamefont {C.~J.}\ \bibnamefont
  {P\'erez~Vicente}}, \bibinfo {author} {\bibfnamefont {F.}~\bibnamefont
  {Ritort}}, \ and\ \bibinfo {author} {\bibfnamefont {R.}~\bibnamefont
  {Spigler}},\ }\href {\doibase 10.1103/RevModPhys.77.137} {\bibfield
  {journal} {\bibinfo  {journal} {Rev. Mod. Phys.}\ }\textbf {\bibinfo {volume}
  {77}},\ \bibinfo {pages} {137} (\bibinfo {year} {2005})}\BibitemShut
  {NoStop}%
\bibitem [{\citenamefont {Rodrigues}\ \emph {et~al.}(2016)\citenamefont
  {Rodrigues}, \citenamefont {Peron}, \citenamefont {Ji},\ and\ \citenamefont
  {Kurths}}]{Rodrigues/etal:2016}%
  \BibitemOpen
  \bibfield  {author} {\bibinfo {author} {\bibfnamefont {F.~A.}\ \bibnamefont
  {Rodrigues}}, \bibinfo {author} {\bibfnamefont {T.~K.~D.}\ \bibnamefont
  {Peron}}, \bibinfo {author} {\bibfnamefont {P.}~\bibnamefont {Ji}}, \ and\
  \bibinfo {author} {\bibfnamefont {J.}~\bibnamefont {Kurths}},\ }\href
  {\doibase 10.1016/j.physrep.2015.10.008} {\bibfield  {journal} {\bibinfo
  {journal} {Phys. Rep.}\ }\textbf {\bibinfo {volume} {610}},\ \bibinfo {pages}
  {1} (\bibinfo {year} {2016})}\BibitemShut {NoStop}%
\bibitem [{\citenamefont {Motter}\ \emph {et~al.}(2013)\citenamefont {Motter},
  \citenamefont {Myers}, \citenamefont {Anghel},\ and\ \citenamefont
  {Nishikawa}}]{Motter/etal:2013}%
  \BibitemOpen
  \bibfield  {author} {\bibinfo {author} {\bibfnamefont {A.~E.}\ \bibnamefont
  {Motter}}, \bibinfo {author} {\bibfnamefont {S.~A.}\ \bibnamefont {Myers}},
  \bibinfo {author} {\bibfnamefont {M.}~\bibnamefont {Anghel}}, \ and\ \bibinfo
  {author} {\bibfnamefont {T.}~\bibnamefont {Nishikawa}},\ }\href
  {https://doi.org/10.1038/nphys2535} {\bibfield  {journal} {\bibinfo
  {journal} {Nat. Phys.}\ }\textbf {\bibinfo {volume} {9}},\ \bibinfo {pages}
  {191 EP } (\bibinfo {year} {2013})}\BibitemShut {NoStop}%
\bibitem [{\citenamefont {Trias}(2012)}]{Trias:2012}%
  \BibitemOpen
  \bibfield  {author} {\bibinfo {author} {\bibfnamefont {A.}~\bibnamefont
  {Trias}},\ }in\ \href {\doibase 10.1109/PESGM.2012.6344759} {\emph {\bibinfo
  {booktitle} {IEEE Power and Energy Society General Meeting}}}\ (\bibinfo
  {year} {2012})\ pp.\ \bibinfo {pages} {1--8}\BibitemShut {NoStop}%
\bibitem [{\citenamefont {Milan}\ \emph {et~al.}(2014)\citenamefont {Milan},
  \citenamefont {W\"achter},\ and\ \citenamefont {Peinke}}]{Milan/etal:2014}%
  \BibitemOpen
  \bibfield  {author} {\bibinfo {author} {\bibfnamefont {P.}~\bibnamefont
  {Milan}}, \bibinfo {author} {\bibfnamefont {M.}~\bibnamefont {W\"achter}}, \
  and\ \bibinfo {author} {\bibfnamefont {J.}~\bibnamefont {Peinke}},\
  }\href@noop {} {\bibfield  {journal} {\bibinfo  {journal} {J. Renewable
  Sustainable Energy}\ }\textbf {\bibinfo {volume} {6}},\ \bibinfo {pages}
  {033119} (\bibinfo {year} {2014})}\BibitemShut {NoStop}%
\bibitem [{\citenamefont {Manwell}\ \emph {et~al.}(2009)\citenamefont
  {Manwell}, \citenamefont {McGowan},\ and\ \citenamefont
  {Rogers}}]{Manwell:2009}%
  \BibitemOpen
  \bibfield  {author} {\bibinfo {author} {\bibfnamefont {J.~F.}\ \bibnamefont
  {Manwell}}, \bibinfo {author} {\bibfnamefont {J.~G.}\ \bibnamefont
  {McGowan}}, \ and\ \bibinfo {author} {\bibfnamefont {A.~L.}\ \bibnamefont
  {Rogers}},\ }\href@noop {} {\emph {\bibinfo {title} {Wind energy
  explained}}}\ (\bibinfo  {publisher} {John Wiley \& Sons, Ltd},\ \bibinfo
  {year} {2009})\BibitemShut {NoStop}%
\bibitem [{\citenamefont {B{\"o}ttcher}\ \emph {et~al.}()\citenamefont
  {B{\"o}ttcher}, \citenamefont {Otto}, \citenamefont {Kettemann},\ and\
  \citenamefont {Agert}}]{Boettcher/etal:2019}%
  \BibitemOpen
  \bibfield  {author} {\bibinfo {author} {\bibfnamefont {P.~C.}\ \bibnamefont
  {B{\"o}ttcher}}, \bibinfo {author} {\bibfnamefont {A.}~\bibnamefont {Otto}},
  \bibinfo {author} {\bibfnamefont {S.}~\bibnamefont {Kettemann}}, \ and\
  \bibinfo {author} {\bibfnamefont {C.}~\bibnamefont {Agert}},\ }\href@noop {}
  {\enquote {\bibinfo {title} {Time delay effects in the control of synchronous
  electricity grids},}\ }\bibinfo {howpublished} {arXiv:1907.13370
  (2019)}\BibitemShut {NoStop}%
\bibitem [{Note3()}]{Note3}%
  \BibitemOpen
  \bibinfo {note} {M.\ Brune, ``Modeling of wind speeds with the Langevin
  equation'' (in German), bachelor thesis (2018), Osnabr{\"u}ck
  University.}\BibitemShut {Stop}%
\end{thebibliography}


\end{document}